\documentclass[12pt,preprint]{aastex}
\shorttitle{RELATIVISTIC PHASE SHIFT IN PULSARS}
\shortauthors{R. T. Gangadhara}
\received{}
\input epsf.tex
\usepackage{graphicx}

\begin{document}
\title{MILLISECOND PULSAR EMISSION ALTITUDE FROM RELATIVISTIC PHASE SHIFT: PSR~J0437-4715}

\author{R. T. Gangadhara and R. M. C. Thomas}
\affil{Indian Institute of Astrophysics, Bangalore -- 560034, India\\
\email{ganga@iiap.res.in}}

\begin{abstract}
We have analyzed the profile of the millisecond pulsar PSR~J0437-4715 at 1440 MHz by fitting the 
Gaussians to pulse components, and identified its 11 emission components. We propose that 
they form a emission beam with 5 nested cones centered on the core. Using the phase location 
of component peaks, we have estimated the aberration--retardation (A/R) phase shift.
Due to  A/R phase shift, the centroid of intensity profile and 
the inflection point of polarization angle swing are symmetrically shifted in the opposite directions
with respect to the meridional plane, which is defined by the rotation and magnetic axes. By 
recognizing this fact, we have been able to locate the
phase location of meridional plane and estimate the absolute altitude of emission of core and conal components
relative to the neutron star center. Using the more exact expression for phase shift given recently by
Gangadhara (2005),  we find  that the radio emission comes from a range of altitude
starting from the core at 7\% of light cylinder radius to outer most cone at 30\%.
\end{abstract}
\keywords{pulsars: general ---pulsars: individual (PSR~J0437-4715)}

\section{INTRODUCTION}
\label{sec1}
The shape and polarization of integrated pulse of a pulsar are quite stable, and they are 
expected to the reflect the magnetic field geometry and the structure of pulsar emission beam. 
The common occurrence of
odd number of components and their distribution across the pulse window has led to the
nomenclature of core and cone structure for the pulsar emission beam (Rankin 1983a).
However, Lyne \& Manchester (1988) and Manchester (1995) have argued for a different 
interpretation, in which components on either side of core arise from a random
distribution of emitting regions, which lead to the so-called patchy cone.
Mitra \& Deshpande (1999) have attempted to test these models, and found the evidences
in favor of conal structure.
The frequency evolution of profile structure and polarization angle swing of pulsars 
has been interpreted by considering the radio emission by relativistic plasma
moving in dipolar magnetic field lines
(e.g., Radhakrishan \& Cooke 1969;  Sturrock 1971; Ruderman \& Sutherland 1975;
Rankin 1983a\&b, 1990, 1993; Lyne \& Manchester 1988; Blaskiewicz et~al. 1991; 
Hibschman \& Arons 2001; Mitra \& Li 2004).  

For understanding the pulsar radio emission mechanism, the radio emission altitude
bears an at most importance. There have been two  kinds of methods proposed
for estimating the radio emission altitudes: (1) {\it Purely geometric 
method,} which assumes the pulse edge is emitted from the last open field lines (e.g., Cordes 1978; 
Gil \& Kijak 1993; Kijak \& Gil 2003), (2) {\it Relativistic phase shift method,} which 
assumes the asymmetry in the conal components phase location, relative to core, is due to the 
aberration-retardation phase shift (e.g., Gangadhara \& Gupta 2001, hereafter GG01).
Both the methods have merits and demerits: former method has an ambiguity in identifying
the last open field lines while the later rely on the clear identification of core-cone structure 
of the emission beam. Thomas \& Gangadhara (2005) have showed that the profiles becomes asymmetric
due to the influence of rotation on the radio emitting sources moving along the dipolar magnetic field
lines.

By assuming the pulsar radio emission comes from fixed altitude, Blaskiewicz et~al.~(1991) 
have showed that polarization angle inflection point (PAIP) lags with respect to the centroid 
of the intensity profile by $\sim 4 r/r_{\rm LC},$ where $r$ is the emission
altitude measured from the center of neutron star and $r_{\rm LC}=P c/2 \pi$ is the light cylinder radius.
The parameters $P$ and $c$ are the pulsar spin period and speed of light, respectively.
The results of purely geometric method are found to be in rough agreement with those of 
Blaskiewicz et~al.~(1991). However, the relativistic phase shift method
clearly indicates that the emission altitude across the pulse window is not constant (GG01; 
Gupta \& Gangadhara 2003, hereafter GG03; Johnston \& Weisberg 2006). Dyks, Rudak and Harding (2004, 
hereafter DRH04) by revising the 
relation for aberration phase shift given in GG01, have re-estimated the 
emission heights. The average emission heights are found to be comparable with those of geometric method.

By considering the relativistically beamed radio emission in the direction of magnetic field line 
tangents, Gangadhara (2004, hereafter G04) has solved the viewing geometry in an inclined and 
rotating dipole magnetic field. He has showed that due to geometric restrictions in receiving the beamed 
emission, a distant observer can not receive the radio waves from a constant altitude.
A more exact expression for relativistic phase shift (A/R) is given by Gangadhara (2005; hereafter G05), 
which includes the phase shift due to polar cap currents also. Under small angle approximations, the 
expression of G05 reduces to that of DRH04. 

Recently, Johnston \& Weisberg (2006) have estimated the emission heights of the young pulsar 
PSR J1015-5719 using the expression for A/R phase shift given in G05. In this paper, we intend to estimate
the emission heights of the nearby, bright millisecond pulsar PSR~J0437-4715. It was discovered in the 
Parkes southern survey (Johnston et~al. 1993; Manchester et~al. 1996), has a pulse period of 5.75 ms.
Navarro et~al. (1997) have presented the mean pulse observations of PSR~J0437-4715 at multiple radio 
frequencies, and found a frequency dependent lag of the total-intensity profile with respect to the 
polarization profile.  Jenet et~al. (1998) have
presented the results based on the high time resolution single pulse observations of this pulsar. They 
find the single pulse properties of PSR~J0437-4715 are similar to those of the common slow-rotating pulsars.

   The outline of this paper is as follows. In \S2, we discuss a method for identifying
the phase location of meridional plane. Next, in \S3 we consider the mean pulse data of millisecond 
PSR~J0437-4715, and fit the Gaussian to pulse components. By assuming conal emission beam, we have 
estimated the A/R phase shifts and the emission heights.

\section{Phase of meridional plane}
In the nested cone structure of pulsar emission beam,  the conal components are expected to be
distributed symmetrically on either side of core when observed in the corotating frame. However,
for an observer in the laboratory frame the profile becomes asymmetric due to rotation. A typical 
asymmetric profile is given in Figure~1, and such an asymmetry in the real profiles has 
been reported for many pulsars (e.g., GG01; GG03; Johnston \& Weisberg 2006). 

Blaskiewicz et~al. (1991) have generalized the rotating vector model (RVM) by considering the
first order relativistic effects, which arise due to the pulsar rotation. By assuming that the
emission across the entire pulse comes from a constant height, they have showed that the centroid 
of the total intensity pulse advances to earlier phase by $\sim r/r_{\rm LC}$ while the polarization angle
inflection point (PAIP) is delayed to later phase by $\sim 3\, r/r_{\rm LC},$ where $r$
is the emission height measured from the center of neutron star. However, there are clear evidences
to show that the emission do not come from a fixed altitude (GG01; GG03; Johnston \& Weisberg 2006). 
Consider the intensity profile given in Figure~1, in which, $C_l$ and $C_t$ are the components of an arbitrary cone with center at CC. If the emission altitude is not constant 
across the pulse window then the retardation also  contributes to the phase shift, and it 
is comparable to that of aberration (G05).  So, due to A/R effects, the centroid 
intensity profile is advanced to earlier phase by $\sim 2\, r/r_{\rm LC}$  
while the corresponding PAIP is delayed to later phase by $\sim 2 \,r/r_{\rm LC}$ (e.g., DRH04; Fung 2005). 
Hence, the cone center CC and PAIP lie symmetrically on either side of meridional plane M, which is at the 
phase $\phi'=0^\circ.$ The meridional plane is defined by the plane containing the rotation axis and 
the magnetic axis. Note that the phase location of M is invariant with respect to $r.$
In other words, both CC and PAIP come closer to M at smaller $r,$ and move away from it at larger $r.$ 
In the co-rotating frame, i.e., in the absence of rotation, the
A/R effects vanish, and hence core peak and PAIP appear at the meridional plane M.

The field aligned polar cap current do not introduce any significant phase shift into the phase of PAIP
but it does introduce a positive offset into polarization angle, however, it roughly cancels due to the 
negative offset by  aberration (Hibschman \& Arons 2001). The phase shift of pulse components due to
polar cap current is estimated recently by G05, and found to be quite small compared
to A/R phase shift. 

\section{Emission height in millisecond pulsar: PSR J0437-4715 }
  Consider the nearest bright millisecond pulsar PSR J0457-4715, which has a period of 5.75~ms. 
It was discovered in Parkes southern survey by Johnston et~al. (1993), and is in a binary system 
with white dwarf companion. Manchester \& Johnston (1995) have presented the mean pulse 
polarization properties of PSR J0437-4715 at 1440 MHz. There is significant linear and circular 
polarization across the pulse with rapid changes near the pulse center. The position angle has 
a complex swing across the pulse, which is not well fitted with the rotating vector model, 
probably, due to the presence of orthogonal polarization modes. The mean intensity pulse has a 
strong peak near the center, where the circular polarization shows a clear sense reversal and 
the polarization angle has a rapid sweep. These two features strongly indicate that it is a core. 
The profile shows more than 8 identifiable components. 
\begin{deluxetable}{crrrrrrccl}
\tabletypesize{\scriptsize}
\tablecaption{Parameters related to radio emission from PSR~J0437$-$4715 at 1440 MHz\label{tab2}}
\tablewidth{0pt}
\tablehead{\colhead{Cone No.} & \colhead{$\phi'_{\rm l}$} & \colhead{$\phi'_{\rm t}$} &
\colhead{$\delta\phi'$} & \colhead{$\Gamma$}& \multicolumn{3}{c}{$r$ } & \colhead{$\Delta$}  &
\colhead{$s/s_{_{\rm lof}}$}\\
 &\colhead{$(^\circ)$} & \colhead{$(^\circ)$} & \colhead{$(^\circ)$} & \colhead{$(^\circ)$} & 
\colhead{(Km)\tablenotemark{a}} & \colhead{(Km)\tablenotemark{b}}&\colhead{(\%$r_{\rm LC}$)}& \colhead{(\%)}  & \\
\colhead{(1)}   &  \colhead{(2)}  &  \colhead{(3)}      &           \colhead{(4)}          
&  \colhead{(5)}                 & \colhead{(6)}        
      &        \colhead{(7)}        &   \colhead{(8)}              & \colhead{(9)}          & \colhead{ (10)} }\\
\startdata
0    &   ---~~~~~~    &    ---~~~~~~    &    $-$9.50$\pm$0.26  &    4.00$\pm$0.00  & 22.8    
     &    20.3$\pm$0.6    &    7  &    12    &    0.17$\pm$0.00 \\    
1    &    $-$29.50$\pm$0.24    &    8.50$\pm$0.46    &    $-$10.50$\pm$0.26    &    
          7.06$\pm$0.06    &    25.2    &    23.3$\pm$0.5    &    9  &    8  
      &    0.28$\pm$0.01\\    
2    &    $-$50.00$\pm$0.50    &    21.50$\pm$0.54    &    $-$14.25$\pm$0.37    &    11.50$\pm$0.10    &    
    34.2    &    34.3$\pm$0.9    &    13  &    0    &    0.38$\pm$0.01\\    
3    &    $-$81.50$\pm$0.15   &    37.94$\pm$0.08    &    $-$21.78$\pm$0.08    &    18.00$\pm$0.02    &    
      52.2  &    64.3$\pm$0.3    &    23  &    19    &    0.43$\pm$0.00\\    
4    &    $-$102.01$\pm$0.36    &    56.14$\pm$0.29   
     &    $-$22.90$\pm$0.23    &    22.93$\pm$0.05    &    55.0    &    
    84.9$\pm$0.9    &    31  &    35    &    0.47$\pm$0.00\\    
5   &    $-$129.50$\pm$1.01 &  89.54$\pm$4.15  & $-$19.98$\pm$2.10  &  29.30$\pm$0.40 &  48.0    &    
      91.8$\pm$6.2 &  33  &    49 & 0.57$\pm$0.02\\
\enddata
\tablenotetext{a}{Emission heights computed from the approximate 
formula of DRH04, and $^b$those using the more exact formula of G05.}
\end{deluxetable}
Consider the average pulse profile given in Figure~\ref{profile} for PSR~J0437-4715 at 1440 MHz. 
To identify pulse components and to estimate their peak locations we followed the procedure 
of Gaussian fitting to pulse components. Two approaches have been developed and used by different
authors. Unlike Kramer et~al. (1994), who fitted the {\it sum} of Gaussians to the total pulse
profile, we separately fitted a single Gaussian to each of the pulse components.
We used the package Statistics`NonlinearFit` in Mathematica
for fitting Gaussians to pulse components. It can fit the data to the model with the named
variables and parameters, return the model evaluated at the parameter estimates achieving the
least­squares fit. The steps are as follows:
(i) Fitted a Gaussian to the core component (VI),
(ii) Subtracted the core fitted Gaussian from the data (raw),
(iii) The residual data was then fitted for the next strongest peak, i.e., the component IV,
(iv) Added the two Gaussians and subtracted from the raw data,
(v) Next, fitted a Gaussian to the strongest peak (IX) in the residual data.
   The procedure was repeated for other peaks till the residual data has no prominent
peak above the off pulse noise level.

By this procedure we have been able to identify 11 of its emission components: I, II, III, IV, 
V, VI, VII, VIII, IX, X and XI, as indicated by the 11 Gaussians in Figure~\ref{profile}. The 
distribution of conal components about the core (VI) reflect the core-cone structure of the 
emission beam. We propose that they can be paired into 5 nested cones with core  at the center. 

Manchester \& Johnston (1995) have fitted the rotating vector model (RVM)
of Radhakrishnan \& Cooke (1969) to polarization angle data of PSR J0437-4715 and
estimated the polarization parameters $\alpha=145^\circ$ and 
$\zeta=140^\circ.$ The parameter $\alpha$ is the magnetic axis inclination angle relative to the
pulsar spin axis, and $\zeta=\alpha+\beta ,$
where $\beta$ is the sight line impact angle relative to the magnetic axis.
 However, for these values of $\alpha$ and $\zeta$ 
the colatitude $s/s_{\rm lof}$ (see eq.~[15] in GG01) exceeds 1 for all the cones.
The parameter $s$ gives the foot location of field lines, which are associated with the conal 
emissions, relative to  magnetic axis on the polar cap.  It is normalized with the colatitude 
$s_{\rm lof}$ of last open field line.  However, Gil and Krawczyk (1997) have reported
$\alpha=20^\circ$ and $\beta=-4^\circ$ by fitting the average intensity pulse profile 
rather than by formally fitting the RVM. Further, they find the 
relativistic RVM by Blaskiewicz et~al. (1991) calculated with 
these $\alpha$ and $\beta$ seems to fit the observed position angle quite well than the 
non-relativistic RVM.  Further, Gil and Krawczyk (1997) have reported that the centroid of
intensity profile and PAIP are separated by $\sim 40^\circ.$ By taking the cone center (CC) of outermost
cone as the centroid of intensity profile, we can locate the phase of meridional plane M. It is at the
mid point, i.e., at $\delta\phi'\sim 20^\circ$ between CC and PAIP. Using $\delta\phi'\approx 2 r/r_{\rm LC},$
we get the emission height $r\approx 48$~km. We have adopted these values of $\alpha$ and $\beta$ in our model 
as they also confirm $s/s_{\rm lof}\leq 1$ (see below). The phase $\phi'=0^\circ,$ in Figure~2, corresponds
to the phase of meridional plane (M), which is at the mid point between CC of outermost cone and PAIP of relativistic
polarization angle swing.

In column~1 of Table~\ref{tab2} we have given the cone numbers. The cone number 0 stands for core component.  
The peak locations of conal components on leading and trailing sides are given in columns~2 and 3, respectively.
The conal components are believed to arise from the nested hollow cones of emission
(Rankin 1983a, b, 1990, 1993), which along with the central core emission, make up the pulsar
emission beam.  Let $\phi'_l$ and $\phi'_t$ be the peak locations of conal components on leading
and trailing sides of a pulse profile, respectively. Then,  using the following
equations, we estimate the phase shift $(\delta\phi')$ of cone center with respect to M and the phase
location $(\phi')$ of component peaks in the absence of phase shift, i.e., in the corotating frame 
(see eq.~[11] in GG01):
\begin{equation}\label{dpphip}
\delta\phi'=\frac{1}{2}(\phi'_t+\phi'_l)~,\quad\quad
\phi'=\frac{1}{2}(\phi'_t-\phi'_l)~.
\end{equation}
In column~4 of Table~1, we have given the values of $\delta\phi'.$ It increases in magnitude from inner most cone
to 4th cone, but for outermost cone i.e., for 5th cone it decreases slightly. We suspect this decrease in $\delta\phi'$ 
for the 5th cone is due to the fact that emission spot moves closer to the rotation axis, and hence aberration
phase shift becomes smaller (G05).

The half-opening angle $\Gamma$ (see eq.~[7] in G04) of the emission beam is given in column~5, and it is about 
$30^\circ$ for the last cone. In column~6, we have given the emission heights computed from the approximate 
expression given in DRH04. Using the more exact formalism for A/R phase shift given in G05 (see eq.~[45]),
we computed the emission heights and given in column~7. Their percentage values in $r_{\rm LC}$ are given in
column~8. It shows the emission in PSR~J0437-4715 occurs over a range of altitude starting from the core at 20 Km to 
the outermost cone at 90 Km. In coulmn~9, we have given the refinement $\Delta(\%)=(1-c_6/c_7) 100,$ where $c_6$ and 
$c_7$ are the emission heights given in columns~6 and 7, respectively. It shows the refinement is
quite significant for the outer cones. In column~10, we have given the colatitude $s/s_{\rm lof}$ 
of foot field lines,  which are associated with the component emissions. It shows due to the relativistic beaming
and geometric restrictions, observer tends to receive the emissions from the open field lines which are 
located in the colatitude range of 0.2 to 0.6 on the polar cap.
\section{DISCUSSION}
A long standing question in pulsar astronomy has been the location of the radio emission region
in magnetosphere. The emission heights of PSR B0329+54 given in GG01, six other pulsars in GG03
and the revised ones in DRH04 are all relative to the emission height of core.
They assumed that the core emission height is negligible compared to the light cylinder radius of 
classical pulsars, but it may not be so in the case of millisecond pulsars.
Though Blaskiewicz et~al. (1991) have assumed a constant emission height in their model, they
have indeed indicated the possibility of core emission at a lower altitude than conal emissions. 
We identify, due to A/R phase shift, that the meridional plane (M) is 
located at the mid point between the phase of core peak and polarization angle inflection 
point (PAIP). By recognizing this fact, we have been able to estimate the {\it absolute} emission heights 
of core as well as conal components.  
We find that the core emission in PSR~J0437$-$4715 is indeed come from a lower altitude 
than the conal emissions. Note that our estimates of error bar do not include the error bars in $\alpha$ and 
$\beta,$  as they are not reported. For comparison we have reported the 
emission heights (see column 6 in Table 1) estimated from the approximate relation of DRH04, which is
independent of the parameters $\alpha$ and $\beta.$ In the case of core and inner cones (1 and 2), 
the refinement in the emission height is $<12$\% (see column 9).

We selected PSR J0437-4715 because of its strong core and clear conal emissions. Polarization angle (PA) is 
complicated near the core component but we can still locate the meridional plane by avoiding 
fitting that part of PA data. 
The idea is that without the knowledge of polarization angle data near the core it is
possible to locate the meridional plane by just fitting the relativistic polarization angle formula of
BCW91 to segments of PA angle data corresponding to the conal components. Since the emission
altitude for each cone is different, one can not fit the entire PA data with a constant altitude.
If we can best fit a segment of PA data corresponding to a outer cone or any other inner cone
it is possible to locate the phase of the meridional plane. Since we used the formula of
Blaskiewicz et al. (1991), in which the aberration phase shift is independent of phase, our
results are approximate.
 
The polarization angle curve obtained by fitting the segments of PA data
under $C_l$ and $C_t$ gives a virtual inflection point (PAIP) corresponding certain height reported by fitting
routine. Since the emission height vary for each cone, we shall have as many number
of virtual inflection points as the cones. The point to be noted is that the cone center and PA virtual
inflection point of each cone is approximately symmetrically located on either side of
the meridional plane. Since the cone centers and the virtual PA inflection points are approximately symmetrically
located on either side of meridional plane, it is possible to locate the phase of
meridional plane even if we are unable to detect virtual PA inflection point for certain cones or core.

The geometrical method, which by comparing the measured pulse widths with geometrical predictions from 
dipolar models, can yield absolute emission heights. But such absolute values are likely to be 
underestimated due to an ambiguity in identifying the last open field 
lines.  The profile widths generally measured at a 10\%-level can give the lower limits for emission 
altitudes (Xilouris et~al. 1996).  In PSR~J0437$-$4715 the components I and XI are at the intensity level
of 1 to 2\%, and the field lines associated with them are not last open field lines, but they are found to be
located at the colatitude $s/s_{\rm lof}\sim 0.6$ on polar cap.

It is possible that the A/R phase shift can be reduced by the rotational distortion
of magnetic field sweep back of the vacuum dipole magnetic field lines. It was first considered in detail 
by Shitov (1983). Dyks and Harding (2004) have investigated the rotational distortion of 
pulsar magnetic field by making the approximation of a vacuum magnetosphere. For $\phi'=110^\circ,$ 
$\beta=-4^\circ$ and $\alpha = 20^\circ$
we computed the phase shift due to magnetic field sweep back $\delta\phi'_{\rm mfsb}$ (see eq. [49] in G05).  
But it is found to be $< 0.6^\circ$ for $r/r_{\rm LC}\leq 0.3, $ which is much
smaller than the aberration, retardation and polar cap current phase shifts in PSR~J0437$-$4715.

\section{CONCLUSION}

We have analyzed the mean profile of the millisecond PSR J0437-4715 at 1440 MHz, and
identified 11 of its emission components. We propose that they form a emission beam consist 
of 5 nested cones  centered on the core. Using the phase location
of component peaks, we have estimated the aberration--retardation (A/R) phase shift.
Due to  A/R phase shift, the center of each cone and the inflection point of corresponding 
polarization angle swing are symmetrically shifted in the opposite directions with respect to 
the meridional plane. By recognizing this fact, we have been able to locate the
phase location of meridional plane and estimate the absolute altitudes of emission of core and 
conal components relative to neutron star center. We used the more exact expression for phase 
shift given recently by G05. The radio emission at 1440 MHz in PSR J0437-4715 is found to come 
from an altitude range starting from the core at 7\% of light cylinder radius to the outer most cone at 30\%.

\begin{acknowledgements}
We used the data available on EPN archive maintained by MPIfR, Bonn. We are thankful to 
M.~N.~Manchester and  S.~Johnston for making their data available on web.
\end{acknowledgements}

\clearpage
\begin{figure}
\vskip -0 truecm
\centerline{ \epsfysize12.5 truecm {\epsffile[127 250 477 667]{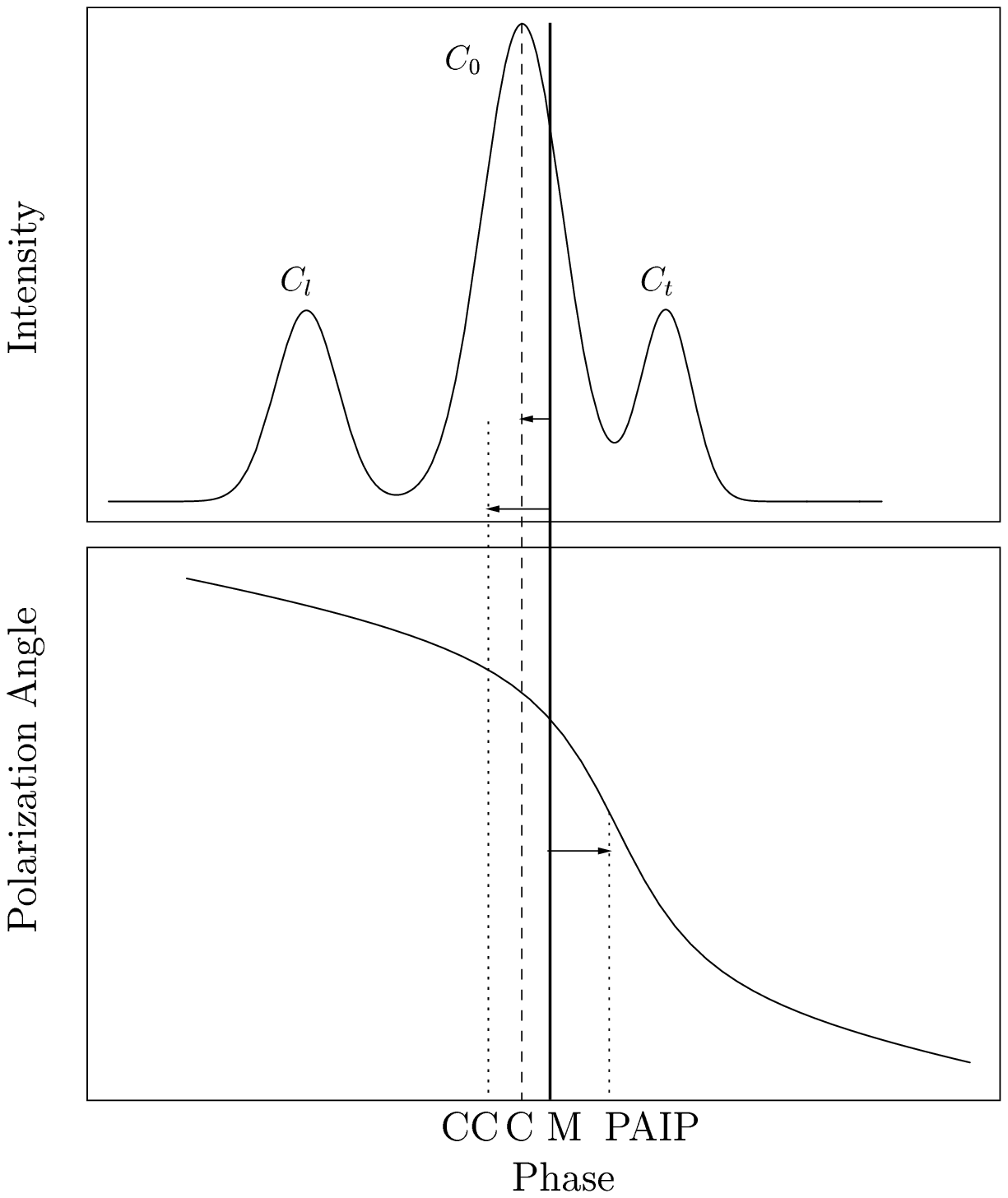}}}
\caption{A typical pulse profile.
$C_0$ is the core. $C_l$ and $C_t$ are the
components on leading and trailing sides, respectively.
The dotted lines represent the phase of cone center (CC) and polarization
angle inflection point (PAIP). 
The thick line mark the phase of meridional plane (M), and the broken line
mark the phase of core peak (C). Arrows indicate direction of phase shift of
CC, C and PAIP with respect to M.}
\label{typical}
\end{figure}

\begin{figure}
\vskip -0 truecm
\hskip 0 truecm \epsfysize10.5 truecm {\epsffile[127 327 655 668]{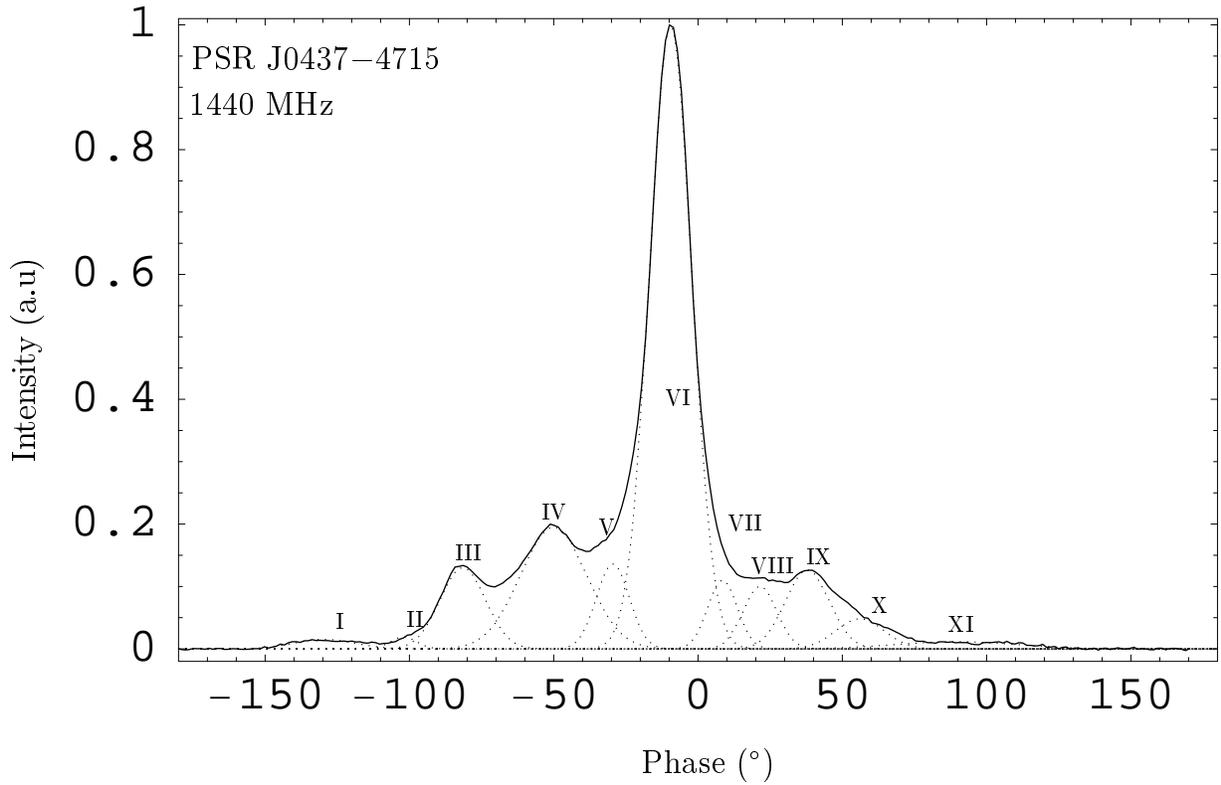}}
\caption{Average pulse of PSR J0437-4715 at 1440 MHz and the model Gaussians (dotted line curves)
fitted to emission components. Due to A/R effects, the core (VI) peak is shifted to
the phase  $\sim -9.5^\circ$ from the meridional plane (M) which is at $0^\circ$.}
\label{profile}
\end{figure}
\end{document}